# When Trade Produces Knowledge: Dynamic Pricing, Bilateral Learning, and Trust

**Chupeng Xie**

## Abstract

A transaction can allocate a product and create information unavailable before consumption. We study dynamic pricing when a pricing agent and a buying agent hold different estimates of match value, traded outcomes update a public belief, and attributed extraction depreciates relationship capital. Public learning polarizes exchange: greater precision drives compatible delegated policies toward trade and incompatible policies toward rejection. The seller's exact expected profit must account for the fact that acceptance selects its posterior margin. Its optimal policy therefore tracks the public margin, trust distance, and posterior precision. Higher extraction reduces both current acceptance and the arrival of future outcome signals, producing a reference-respecting region and an extraction–learning trap. With unequal signal precision, the trading rule additionally selects common quality; a selection-naive learner can become more confident and less accurate. Transactions create an information externality for later participants, while a shared misspecified model can make the agents agree without becoming correct. The analysis distinguishes data production, Bayesian knowledge, relationship capital, and verified computation, and identifies the state and protocol records required to test the mechanisms.

## 1. Introduction

Markets do more than move goods to people who value them. For experience goods, recommendations, software, professional services, and many AI-mediated transactions, purchase makes a previously latent match value observable. Before consumption, the buyer, seller, and platform have forecasts. After consumption, they may have a new measurement. The transaction is therefore both an allocation and an experiment.

That distinction matters for algorithmic pricing. A seller that raises the current price loses some current demand. If only current allocation matters, this is the familiar margin–volume trade-off. If consumption produces information, the rejected transactions also remove observations from the future training set. If the same offer is attributed to a persistent seller, aggressive pricing can additionally depreciate trust and make later transactions less likely. A pricing agent that ignores these two state variables can extract more from the customers who buy today while reducing what the market will know tomorrow.

The problem is bilateral. A pricing agent estimates what an item is worth to a particular principal and posts a price. A buying agent separately estimates the principal's match value and applies a delegated acceptance policy. Their estimates can differ even when the agents share the same nominal surplus rule. If trade occurs, an outcome signal becomes available to the buyer and, subject to data permissions, to the platform. The platform aggregates outcomes across transactions and feeds a public posterior into later seller and buyer estimates. Price affects trade; trade affects the production of data; data affect posterior beliefs; beliefs affect later price and acceptance; and attributed pricing affects trust. The resulting loop is

$$\text{pricing} \rightarrow \text{trade} \rightarrow \text{experience signal} \rightarrow \text{posterior update} \rightarrow \text{pricing and trust}. \quad (1)$$

This paper studies that loop. The central dynamic state is not a database count alone. It includes the public margin, posterior precision, and relationship capital. The two novel persistent stocks are knowledge precision and relationship capital. Data become knowledge only through an updating model, and a verified computation can establish which data and rule were used without proving that the maintained model is correct.

The core analysis yields four results. First, public learning **polarizes policy compatibility**. When the seller's restrained schedule leaves more slack than the buyer's mandate requires, greater public precision reduces bilateral disagreement and drives trade probability toward one. When the schedules are incompatible, the same reduction in disagreement drives trade probability toward zero. Better common information does not mechanically produce more exchange; it makes the underlying institutional compatibility decisive.

Second, the seller's dynamic first-order condition contains two terms absent from a static personalized-pricing problem. Raising extraction increases future trust distance, and by reducing the probability of trade it destroys the option value of the post-consumption signal. Under a compatible policy region these terms both reduce optimal extraction. A reference-respecting region arises even when restraint is not imposed as an ethical objective.

Third, pricing can create an **extraction–learning trap**. Relative to a restrained policy, an aggressive policy produces a higher trust state and fewer transactions. Fewer transactions slow the accumulation of public precision; lower precision sustains posterior disagreement and further reduces trade. A coupling result orders the trust, transaction, and knowledge paths. The result differs from a standard exploration–exploitation problem because the seller is not merely choosing an informative action. It is choosing whether the market interaction required to produce the outcome will occur.

Fourth, transaction-generated data are selected data. In the symmetric-signal benchmark, the acceptance statistic is independent of common quality and ordinary Gaussian updating is valid. With unequal signal precision, selection and quality are correlated. We derive the truncated-normal correction and show that ignoring it creates a bias whose magnitude rises as extraction tightens the trading threshold. An attestation can prove that a selection-aware likelihood was executed, but it cannot establish that the likelihood family is correctly specified.

Two extensions define the mechanism's welfare and epistemic boundaries. A transaction can benefit later buyers by improving matching and reducing disagreement. A seller internalizes only the part of this knowledge value that returns as its own future profit. In a two-period welfare benchmark, the privately optimal price is therefore above the socially optimal price whenever the later-buyer information benefit is positive. A per-transaction learning subsidy implements the constrained social price. The sign is not universal: if improved prediction is used primarily for extraction, the net buyer-side information value can be negative. The extension identifies the economic object that must be measured rather than presuming that more data are always socially beneficial.

The second extension shows that agreement is not accuracy. If a high-precision public outcome stream contains a common, unmodeled bias, the pricing and buying agents' estimates converge to one another while both converge to the wrong value. Verification of data provenance and code execution does not remove this common-model risk. External validation or model pluralism is required.

The paper is organized around a decision architecture associated with Herbert Simon. Simon (1977) distinguished intelligence, design, and choice in managerial decision making. We use those categories as an organizing discipline rather than as a behavioral assumption: intelligence is the acquisition and aggregation of transaction signals; design is the selection of the pricing, acceptance, and evidence policies; and choice is their execution by the two agents. We add review as the explicit feedback step in which outcomes update both the public belief and the trust state. The formal model therefore gives the decision cycle an economic state transition rather than treating it as a descriptive checklist.

The closest literatures study important pieces of the loop separately. Bayesian consumer-learning models show how usage and communication change purchase beliefs (Erdem and Keane 1996; Villas-Boas 2004; Narayanan and Manchanda 2009; Zhao et al. 2013). Strategic-learning models make prices affect experimentation (Bergemann and Valimaki 1996, 2000), while recent pricing work studies demand learning, reference effects, and adaptive price exploration (den Boer and Zwart 2014; den Boer and Keskin 2022; Jain et al. 2024). Data-economics papers analyze cross-user learning and externalities (Ichihashi 2021; Hagiu and Wright 2023). Research on personalized prices and fairness shows why the same information that improves targeting can provoke costly responses (Anderson and Simester 2010; Allender et al. 2021; Dube and Misra 2023). Our increment is to join bilateral posterior disagreement, trade-contingent information production, an endogenous trust stock, and a verifiable updating boundary in a single dynamic pricing problem.

This integration changes the object of empirical analysis. A sale/no-sale panel is insufficient. Identification requires both agents' information summaries or committed score bins, the offer and acceptance rule, whether an outcome signal was produced, the version of the updating model, the public posterior before and after the outcome, and the persistent relationship identifier. Those fields are increasingly feasible in controlled agent markets and protocol-mediated commerce, although no existing commercial log is claimed to contain them all.

# 2. Related Literature and Contribution

## 2.1 Consumer learning and experience

Erdem and Keane (1996) model usage experience and advertising as noisy signals about brand attributes and estimate Bayesian learning using scanner data. Villas-Boas (2004) shows that post-purchase learning can create brand loyalty and alter forward-looking price competition. Freimer and Horsky (2008) connect consumption learning to price promotions, while Narayanan and Manchanda (2009) estimate heterogeneous rates of quality learning. Zhao et al. (2013) combine individual experience and online reviews for experiential products. Dzyabura and Hauser (2019) consider consumers who learn their own preference weights through search. These studies establish that consumption can reveal value and that learning affects demand.

Our buyer is not merely learning a fixed brand quality. A buying agent has a private posterior about principal-specific match value and evaluates a price generated from a different seller posterior. The platform then learns a common component only when the bilateral policies permit trade. This distinction makes posterior disagreement and mandate compatibility part of the data-generating process.

## 2.2 Strategic experimentation and dynamic pricing

Bergemann and Valimaki (1996) endogenize the price of experimentation when consumers learn product values, and Bergemann and Valimaki (2000) study market experimentation more generally. The revenue-management literature studies firms that learn demand while pricing; controlled-variance pricing, bandit policies, and adaptive exploration balance current revenue against parameter learning (den Boer and Zwart 2014; den Boer and Keskin 2022; Jain et al. 2024). Atasu, Ciocan, and Desir (2024) study delegation when a sales agent learns valuations and prices. Feng, Li, and Zhang (2019) analyze dynamic pricing in response to online reviews.

In those settings the learning state normally belongs to a consumer, a firm, or a public review process. Here, the seller and buyer have distinct private estimates, and the outcome that updates the public state is itself gated by their compatibility. Moreover, extraction affects a second persistent state–trust–that changes future acceptance even conditional on posterior precision. The seller thus faces a three-way margin among current extraction, relationship capital, and knowledge production.

### 2.3 Data-enabled learning and information externalities

Transaction records can be predictive of other users. Ichihashi (2021) studies data externalities when one consumer's data reveal information about others. Hagiu and Wright (2023) distinguish across-user and within-user data-enabled learning and show how learning shapes competitive advantage. Bergemann, Bonatti, and Gan (2020) analyze the social dimension of individual data. This literature makes clear that the value created by one record need not accrue to the person who generated it.

We endogenize whether the record exists. A platform cannot observe the paper's experience signal if the price prevents consumption. The seller's price is therefore also a data-acquisition decision. The welfare analysis separates the seller's continuation value from the value that improved public beliefs create for later buyers.

### 2.4 Trust, fairness, and verifiable rules

Kahneman, Knetsch, and Thaler (1986) document reference-based judgments of price fairness. Anderson and Simester (2010) show in a field experiment that perceived price unfairness can depress later demand, especially among valuable customers. Allender et al. (2021) model and experimentally study fairness concerns and strategic price obfuscation. Dube and Misra (2023) show that machine-learning personalization can raise profit while redistributing surplus. These findings motivate a persistent trust state rather than an assumption that every pricing encounter begins afresh.

The verification literature asks whether a rule, input, or computation can be credibly established. Our use is deliberately narrow. A certificate may establish provenance, policy version, and execution of a specified Bayesian or selection-corrected update. It does not reveal the latent true value and cannot prove that an omitted common disturbance equals zero. This evidence boundary is central to the shared-model-risk result.

### 2.5 Boundary relative to demand learning and algorithmic pricing

It is useful to state what the paper is not doing. In classical demand learning, the unknown object is commonly a demand curve or one of its parameters. A seller varies prices, observes purchase indicators, and learns which price maximizes revenue. Recent bandit methods exploit economic restrictions across prices to learn that curve more efficiently (Weaver, Kumar, and Jain 2025). In the present model the unknown object is a latent match component, and the informative outcome is generated after use. A purchase indicator may reveal something about acceptance, but it is not a substitute for the experience signal. The seller is therefore not choosing a price merely to excite variation in demand; it is choosing whether the interaction that reveals product performance or match quality will happen.

Nor is the paper a model of personalized ranking. Qiu et al. (2025) show that personalized rankings can reduce price elasticity and induce pricing algorithms to charge more even without explicit price discrimination. That mechanism operates through search order. Our mechanism operates after a product has been encountered: the buying agent compares its posterior and mandate with a price based on a seller posterior, and consumption changes the public information state. Ranking can be added upstream by changing which match reaches the bilateral stage, but it does not eliminate the trade-contingent outcome.

The data-rights literature also emphasizes that collection, deletion, portability, and use can have different competitive effects. Cong and Matsushima (2026), for example, show that personal data management can generate cross-market price and welfare effects that are not monotone in the scope of rights. Our welfare extension is consistent with that caution: it assigns no universally positive sign to transaction data and defines later-buyer information value net of extraction, privacy, and error. A larger dataset is not a sufficient statistic for welfare.

Finally, the paper does not assume that a language model intrinsically has seller or buyer preferences. The economic agent is a deployed policy. It becomes a pricing agent when a principal gives it a state, objective, feasible action set, and authority to post prices. It becomes a buying agent when a different principal gives it a valuation signal, an acceptance mandate, and authority to transact. The implementation might be an optimization routine, a rules engine, a learned policy, an LLM with tools, or a hybrid. This policy interpretation is what permits the theory to cover future agentic-commerce systems without making behavioral claims about a particular model family.

The contribution can therefore be located by four differences. First, the information state is shared but the decision posteriors remain bilateral. Second, a transaction gates an outcome experiment rather than only a binary demand observation. Third, price updates a persistent trust state as well as a posterior state. Fourth, verification is modeled at the inferential boundary: it can establish inputs and computation but not latent truth. Removing any one difference returns the model to a familiar special case. Equal seller and buyer information removes mandate incompatibility; exogenous outcomes remove knowledge production from pricing; $\gamma = 0$ removes relationship capital; and a known, correctly specified likelihood removes the distinction between verified execution and validated inference.

# 3. Environment

## 3.1 Timing and decision architecture

Time is discrete. A platform supports repeated transactions between a seller organization and a sequence or panel of buyer principals. A pricing agent acts for the seller; a buying agent acts for the current buyer. Each period has four economically distinct phases:

1. **Intelligence.** The platform supplies a public prior; each agent obtains a private signal.
2. **Design.** The seller's principal has selected a pricing schedule and the buyer's principal has selected an acceptance mandate. The pricing agent chooses the permitted extraction adjustment.
3. **Choice.** The seller posts a price and the buying agent accepts or rejects.
4. **Review.** If trade occurs, consumption generates an outcome signal. The platform updates the public posterior, and the attributable offer updates relationship trust.

The first three labels adapt Simon's decision phases. Review makes the intertemporal feedback explicit. Human principals choose the policy class ex ante; the software agents execute within it. Nothing in the model requires an LLM. An agent is a policy mapping from its information and mandate into an action.

### 3.2 Latent value and bilateral signals

The product has a common latent component

$$\theta / P_t \sim N\left(\mu_t, \tau_t^{-1}\right), \tag{2}$$

where $P_t = \left(\mu_t, \tau_t\right)$ is the public posterior at the beginning of period $t$. The current principal's match value can additionally contain a mean-zero idiosyncratic component. To keep the bilateral compatibility result transparent, that component is absorbed into the private signal noise in the baseline.

The pricing and buying agents observe

$$s_t = \theta + \varepsilon_t^S, b_t = \theta + \varepsilon_t^B, \tag{3}$$

with independent Gaussian errors of precision $\tau_S$ and $\tau_B$. Their posterior means are

$$m_t^j = \frac{\tau_t \mu_t + \tau_j q_t^j}{\tau_t + \tau_j}, j \in \{S, B\}, \tag{4}$$

where $q_t^S = s_t$ and $q_t^B = b_t$. The agents can therefore disagree without either making a computational error.

### 3.3 Pricing and acceptance policies

The seller's restrained schedule leaves a surplus cushion $\ell > 0$ relative to its own posterior. The pricing agent chooses extraction $a_t \in [0, \acute{a}]$ and posts

$$p_t = m_t^S - \ell + a_t. \tag{5}$$

The buyer's mandate requires posterior surplus $\underline{u} > 0$ and applies a trust wedge $\chi r_t$, where $r_t \geq 0$ is the public or relationship-specific trust distance. It accepts if

$$p_t \leq m_t^B - \underline{u} - \chi r_t. \tag{6}$$

Let

$$h(a, r) = \ell - \underline{u} - a - \chi r \tag{7}$$

denote institutional slack. Trade occurs precisely when

$$m_t^S - m_t^B \leq h(a_t, r_t). \tag{8}$$

Positive slack means that the two policies are compatible at their common posterior; negative slack means they are incompatible. The buyer's rule is a hard delegated constraint. A richer version may add costly inspection or discretionary review, but neither is needed for the knowledge-production mechanism.

For the main analytical results, set $\tau_S = \tau_B = \kappa$. Conditional on the public posterior, posterior disagreement is normal with mean zero and standard deviation

$$\omega(\tau) = \frac{\sqrt{2\kappa}}{\tau + \kappa}. \tag{9}$$

The probability of trade is therefore

$$D(a, r, \tau) = \Phi\left(\frac{h(a, r)}{\omega(\tau)}\right). \tag{10}$$

### 3.4 Trade-contingent knowledge production

If trade occurs, consumption produces a public-eligible outcome

$$z_t = \theta + v_t, v_t \sim N\left(0, \tau_z^{-1}\right). \tag{11}$$

The signal can represent realized fit, task success, verified return, quality, or another measurable consequence. It is not available when the product is not consumed. With symmetric private-signal precision, the disagreement statistic in equation (8) depends on $\varepsilon_t^S - \varepsilon_t^B$ and is independent of $\theta$. Thus selection into trade is independent of common quality in the baseline and the conjugate update is valid:

$$\tau_{t+1} = \tau_t + \tau_z, \mu_{t+1} = \frac{\tau_t \mu_t + \tau_z z_t}{\tau_t + \tau_z} \text{ if trade occurs}. \tag{12}$$

Without trade, no experience signal is created and $P_{t+1} = P_t$. Unequal signal precision breaks the independence and is studied in Section 7.

### 3.5 Relationship capital and exact seller profit

Extraction above the restrained schedule is attributable to the seller identity and updates trust according to

$$r_{t+1} = min\{\acute{r}, \rho r_t + \gamma a_t\}, 0 \le \rho < 1, \gamma > 0. \tag{13}$$

The update depends on the offer, not on purchase, because a rejected personalized quote can still be observed and attributed. The parameter $\rho$ captures persistence, $\gamma$ the relationship cost of extraction, and $\chi$ its effect on the buyer's mandate. We call $r$ "trust distance" for economic interpretation, but it is not a belief or an emotion of the software. It is a reduced-form relationship-capital state implemented by the principal's policy: past attributed extraction tightens the delegated acceptance rule.

Let marginal cost be $c$, and define the public restrained-schedule margin state as $M_t = \mu_t - \ell - c$. Because the posted price contains $m_t^S$, acceptance selects the seller's realized posterior margin. In the symmetric benchmark,

$$Cov\left(m_t^S, m_t^S - m_t^B / P_t\right) = \frac{\kappa}{\left(\tau_t + \kappa\right)^2} = \frac{\omega\left(\tau_t\right)^2}{2}. \tag{14}$$

Writing $k = h(a, r) / \omega(\tau)$, the exact expected current profit is therefore

$$\pi(M, a, r, \tau) = E\left[\left(m^S - \ell + a - c\right) 1\{G \leq h\} / P\right] = (M + a)\Phi(k) - \frac{\omega(\tau)}{2}\phi(k). \tag{15}$$

The last term is a selected-margin composition effect. Replacing equation (15) by $(M + a)D$ would incorrectly factor two correlated random variables. After a traded outcome, posterior precision becomes $\tau^+ = \tau + \tau_z$, while the public margin follows $M^+ = M + \eta$, where

$$\eta / \tau \sim N\left(0, \frac{\tau_z}{\tau\left(\tau + \tau_z\right)}\right). \tag{16}$$

Thus posterior precision is not a sufficient dynamic state; the public posterior mean, equivalently $M$, must also be retained. The seller discounts by $\beta \in (0, 1)$. For the infinite-horizon computation we use compact state and action sets and project only at their declared numerical boundaries. The analytical identities do not depend on that computational closure.

### 3.6 Markov decision problem and economic interpretation

Conditional on the principals' committed policy class, this is a controlled Markov decision problem, not a strategic Markov-perfect-equilibrium claim. The pricing agent selects a state-dependent extraction schedule, the buying agent mechanically implements equation (6), and the public filter implements equation (12). The sufficient public state is $(M, r, \tau)$. Private signals are integrated out each encounter; traded outcomes move both $M$ and $\tau$.

This formulation makes three commitments explicit. First, the seller does not observe the buyer's private signal before posting. If it did, the seller could condition directly on the buyer cap and bilateral disagreement would become an information-design or mechanism problem. Second, the buyer cannot renegotiate its hard minimum-surplus requirement after observing the offer. This captures delegated purchasing settings in which the principal uses a cap to limit discretion. Third, the public outcome signal is eligible for aggregation. Privacy and consent constraints are introduced later as components of net knowledge value rather than silently ignored.

The price in equation (5) contains two conceptually different parts. The restrained schedule $m^S - \ell$ converts the seller's current estimate into a price while leaving a declared cushion. The action $a$ is discretionary extraction above that schedule. This is more useful than equating a high price with misconduct. A high $m^S$ can justify a high restrained price; the trust update attaches only to the departure from the schedule. Empirically, the decomposition requires the pricing-rule version or a committed benchmark. Observing prices alone cannot identify $a$.

The trust state is also deliberately distinct from the posterior. A bad experience can lower the estimated value of the product, while an aggressive but technically valid offer can raise distrust of the seller. Conflating these states would make it impossible to tell whether demand falls because the object is believed to be worse or because the counterparty is believed to be more extractive. The model's data requirements reflect this separation.

Three familiar benchmarks are nested. If $\tau_z = 0$, transactions do not create public information and equation (18) is a dynamic relationship-pricing model. If $\gamma = \chi = 0$, relationship capital is irrelevant and the seller trades current extraction against the value of future outcomes. If both channels are zero, the optimal action solves the exact static margin–demand problem. These nests will matter for empirical model comparison.

## 4. Public Knowledge and Bilateral Compatibility

The first result isolates what public learning does before introducing the dynamic pricing problem.

**Proposition 1. Knowledge polarization**

Under symmetric private-signal precision, for any fixed $(a, r)$,

$$sign\left(\frac{\partial D(a, r, \tau)}{\partial \tau}\right) = sign(h(a, r)). \tag{17}$$

Moreover, as public precision tends to infinity, $D$ tends to one if $h > 0$, one-half if $h = 0$, and zero if $h < 0$.

The result is not the claim that information destroys trade. It says that common knowledge removes accidental exchange caused by posterior noise. When the delegated policies have positive slack, fewer disagreements recover transactions. When they have negative slack, occasional trade was sustained only because the buyer happened to be more optimistic than the seller; common learning eliminates it.

Figure 1 illustrates the polarization. The knife-edge curve remains at one-half because symmetric disagreement is centered at zero. On either side, knowledge pushes the market toward the outcome encoded by the two mandates.

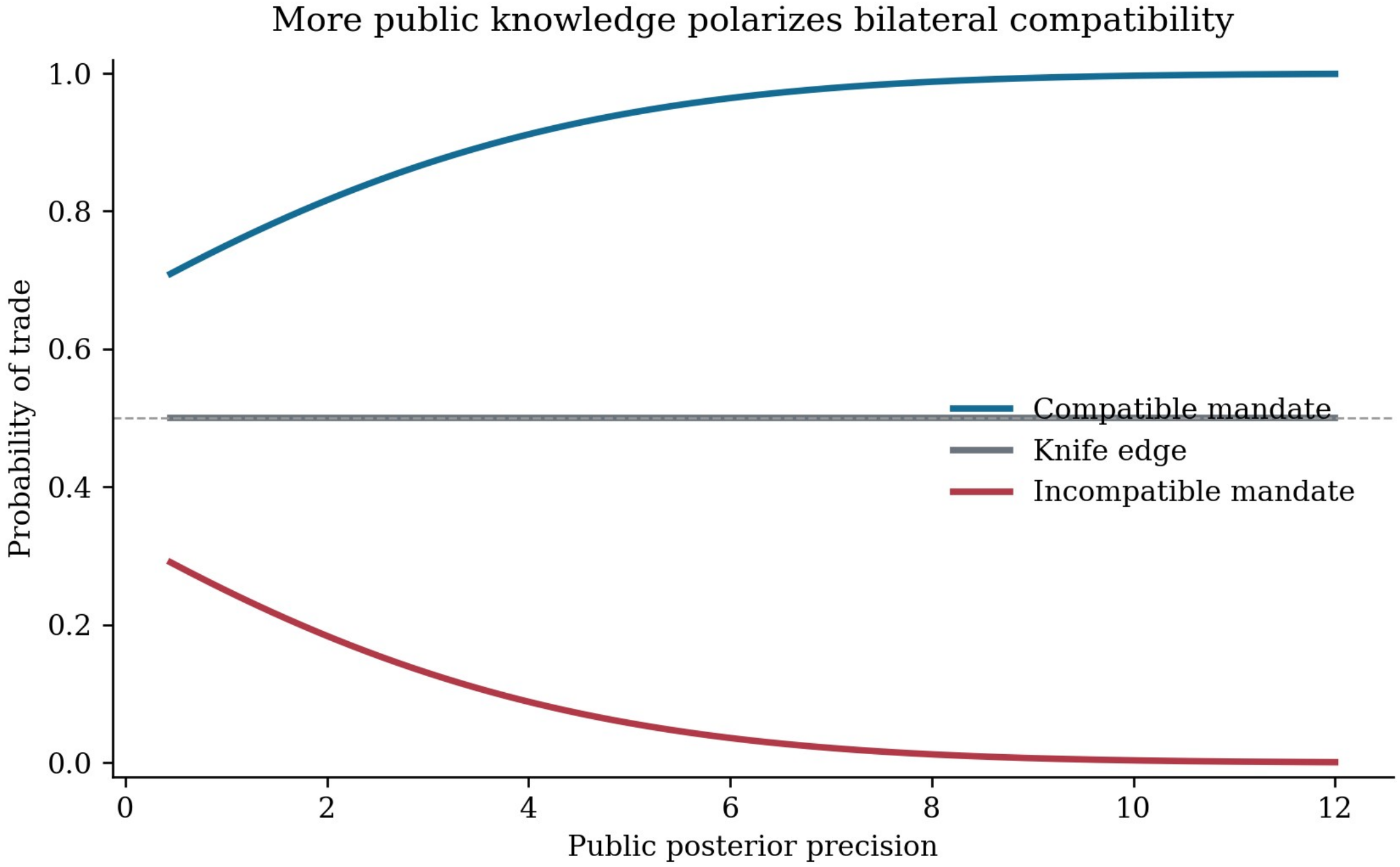

*Figure 1. Public knowledge and bilateral compatibility.*

This result gives a precise interpretation to the claim that trade produces knowledge. The knowledge is not automatically cooperative. It makes the policy architecture more consequential. A platform that improves shared estimates without aligning the seller's and buyer's delegated surplus requirements can make rejection more predictable rather than less frequent.

The comparative statics also distinguish public and private accuracy. Raising $\tau$ makes both agents put more weight on the same public mean and therefore contracts their difference. Raising only $\tau_S$ need not do so: it changes the seller's loading on the common state and can increase the correlation between selection and quality. Similarly, improving the buyer's private measurement can prevent bad purchases while reducing transaction volume. "More accurate agents" is consequently not a complete treatment description. One must state which side receives information, whether it is shared, and how the policy thresholds change.

Knowledge polarization has a protocol-design implication. Suppose a platform observes rising rejection as its common model improves. The model provides two competing explanations. The platform may be learning that the product is a poor match, or it may be revealing a latent incompatibility between the seller's surplus cushion and the buyer's required surplus. A debugging process that examines only prediction error misses the second possibility. Policy slack $h$, not agreement by itself, determines the limiting trade outcome.

The result is also different from adverse selection. Neither side has a privately known fixed type that it strategically signals through trade. Each follows a delegated policy using a noisy estimate. Rejection arises from the difference between those estimates and the two thresholds. Strategic signaling can be added, but it is not required for information to polarize exchange.

## 5. Dynamic Pricing When Trade Produces the Signal

Let $V(M,r,\tau)$ be the seller's value before private signals are realized but after the public state is known. On compact sets $M=[\underline{M},\acute{M}]$, $R=[0,\acute{r}]$, $T=[\underline{\tau},\acute{\tau}]$, and $A=[0,\acute{a}]$, define $M^{+}=\Pi_M(M+\eta)$, $r'=min\{\acute{r},\rho r+\gamma a\}$, and $\tau^{+}=min\{\acute{\tau},\tau+\tau_z\}$. For any candidate action, its complete continuation value is

$$C_V(M,a,r,\tau)=D(a,r,\tau)E_\eta V(M^{+},r',\tau^{+})+[1-D(a,r,\tau)]V(M,r',\tau).$$

The exact Bellman equation is therefore

$$V(M,r,\tau)=\max_{a\in A}\{\pi(M,a,r,\tau)+\beta C_V(M,a,r,\tau)\}. \tag{18}$$

The two continuation states differ because a completed transaction changes both the posterior mean and its precision. The projection $\Pi_M$ is a transparent numerical closure; increasing the margin bounds provides a boundary-robustness check.

Define the expected value of one additional traded outcome at the next relationship state as

$$\Delta_V(M, r', \tau) = E_\eta\left[V\left(M^+, r', \tau^+\right)\right] - V(M, r', \tau). \tag{19}$$

The sign of $\Delta_V$ is an economic condition, not a global consequence of higher precision. Knowledge is valuable in a compatible region when its expected benefit exceeds the effect of selected margins and any approach to the computational boundary. At an interior differentiability point, the exact derivative of the Bellman objective is

$$D + (M + a) D_a - 1/2\, k\, \phi(k) + \beta D_a \Delta_V + \beta \gamma \{ D E_\eta\left[V_r^+\right] + (1 - D) V_r^0 \}. \tag{20}$$

The first three terms are the exact current-profit effect, $\beta D_a \Delta_V$ is the lost-knowledge-flow effect, and the final term is the relationship-capital effect. Here $V_r^+$ and $V_r^0$ evaluate the relationship-state derivative after trade and no trade. The term $-k\, \phi(k)/2$ is the derivative of selected-margin composition. If $\Delta_V \geq 0$ and the relevant relationship derivatives are nonpositive, the last two terms weakly reduce extraction relative to the exact myopic problem.

**Proposition 2. Dynamic restraint and the reference-respecting region**

The Bellman operator in equation (18) is a contraction with modulus $\beta$ and has a unique bounded continuous value function and a nonempty upper-hemicontinuous optimal stationary-policy correspondence. At any state where $\Delta_V \geq 0$ and the expected relationship derivatives in equation (20) are nonpositive, the marginal continuation value of extraction is nonpositive. If the exact state objective is strictly concave in $a$, the dynamic optimum is no more extractive than the exact myopic optimum. The restrained action $a^* = 0$ is optimal whenever the right derivative in equation (20) is nonpositive and the objective is concave on $A$.

The proposition separates two reasons for restraint. The relationship term is familiar from repeated-customer logic: extraction today makes tomorrow's buyer more protective. The knowledge term is distinct. Even if the trust state were fixed, a rejected offer eliminates the outcome experiment and delays public learning. Conversely, even if the platform already knew $\theta$, the relationship term could sustain restraint. The two stocks are complements but not substitutes.

Equation (20) provides a useful accounting identity for a pricing organization. The first bracket is the exact short-horizon return, including who is selected into trade. The second requires a model of how today's purchase changes tomorrow's information set. The third requires persistent identity and a relationship-state model. A firm can give its algorithm access to outcome and reputation data yet omit both continuation effects if its reward contains only current revenue. Data availability does not imply dynamic internalization.

The reference-respecting region should not be read as a universal fixed price. It is a region of states in which the discretionary increment is zero. The underlying restrained price continues to move with $m^S$. The seller can therefore personalize on economically relevant match information while committing not to use the residual flexibility for extraction. This distinction is important for implementation: a protocol can verify a state-dependent schedule rather than force a common list price.

The two continuation costs can reinforce one another. When trust is already high, a marginal increase in $a$ both reduces current acceptance and moves the relationship farther from recovery. The resulting loss of trade also delays the next signal, keeping posterior disagreement high. Conversely, early restraint can start a virtuous path: more trade creates more outcomes, more public precision reduces compatible disagreement, and the higher expected continuation volume raises the value of preserving trust. Proposition 3 formalizes this path ordering under sufficient monotonicity conditions.

There are also states in which the channels conflict. If the mandates are incompatible, Proposition 1 implies that more precision lowers trade. A seller may then privately prefer to delay learning because residual disagreement occasionally lets a transaction through. That incentive is not socially innocuous: it is an incentive to preserve confusion around incompatible policies. The core restraint theorem is therefore stated on a compatible invariant region rather than presented as a global information monotonicity result.

Figure 2 solves the exact three-state Bellman equation. Each panel holds the public margin state fixed. Dark cells are the reference-respecting region. The comparison shows why suppressing $M$ is not innocuous: a seller with a high public margin can optimally choose a different extraction increment even at the same trust and precision state. No global monotonicity in precision is asserted.

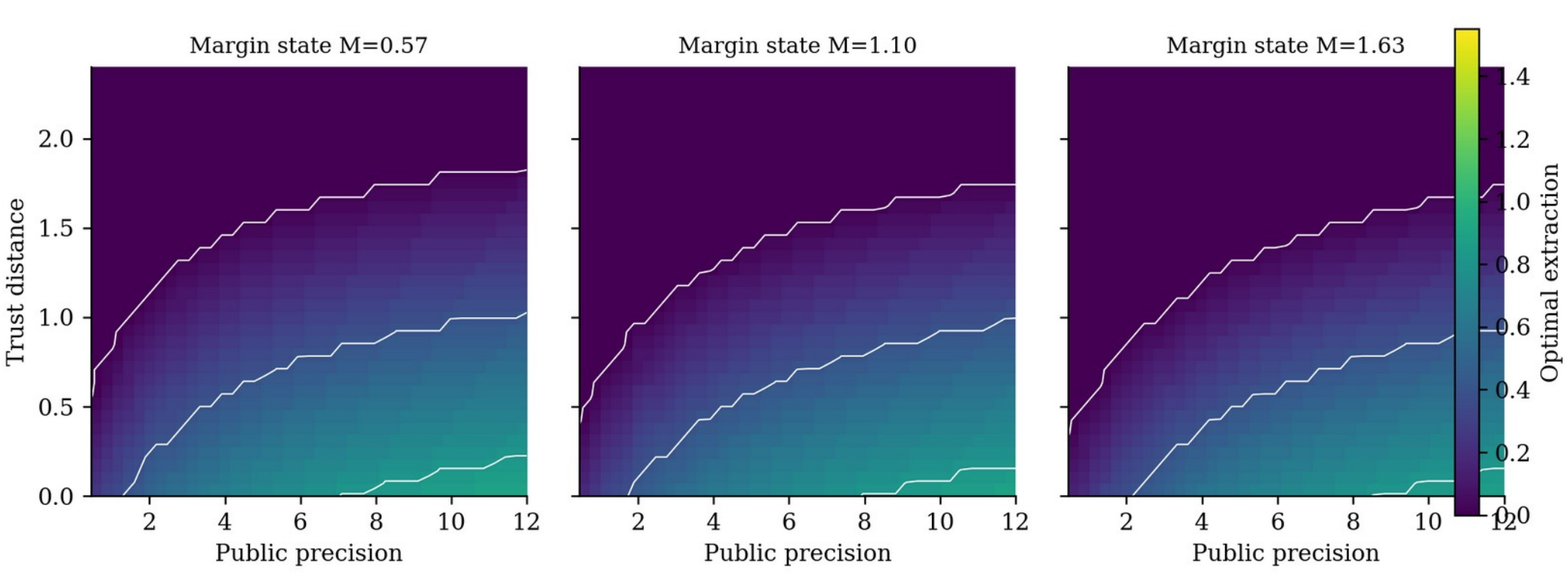

*Figure 2. Optimal extraction over the trust and knowledge states.*

## 6. The Extraction–Learning Trap

A static pricing model treats a rejected offer as a zero-revenue observation. Here it also means that no consumption outcome is produced. Because aggressive offers raise trust distance, they can lower the future arrival rate of observations even after the current price changes.

Consider two Markov policies $A$ and $R$ beginning from the same state. Policy $A$ is everywhere weakly more extractive than $R$. Couple trade outcomes using the same uniform random variable in each period and posterior-mean innovations using the same draw whenever both policies trade.

**Proposition 3. Extraction–learning ordering**

Suppose the coupled paths remain in a region where trade probability is weakly increasing in public precision. If the action ordering $a^A\left(M^A, r^A, \tau^A\right) \geq a^R\left(M^R, r^R, \tau^R\right)$ is preserved on the coupled states, then at every date the aggressive path has weakly higher trust distance, weakly lower public precision, and a weakly smaller cumulative number of transactions, almost surely under the coupling. No ordering of posterior means is required or claimed. Conditional on a fixed current state and action, the expected waiting time for the next outcome signal is $1/D(a, r, \tau)$ and is weakly longer on the aggressive path.

The mechanism is self-reinforcing. Extraction reduces current trade directly. It also raises $r$, which reduces later trade, and it leaves $\tau$ lower, preserving bilateral disagreement. We use “trap” to describe this ordered feedback, not an absorbing state: Gaussian demand remains positive at every finite state.

The condition that trade probability rise with precision is essential. Proposition 1 shows that in an incompatible region greater precision can instead make rejection more certain. In that case, a low-precision state may temporarily sustain accidental transactions. This is another reason not to equate data accumulation with welfare without inspecting the policy slack.

Figure 3 compares the dynamic policy with a myopic policy that maximizes exact profit in equation (15). The plotted transition uses the expected precision increment $\tau_z D$; the public margin remains at its martingale expectation. The Bellman solution itself integrates the full posterior-mean innovation. In this calibration the dynamic seller extracts less early, maintains a smaller trust distance, generates a higher trading rate, and reaches the high-precision region earlier.

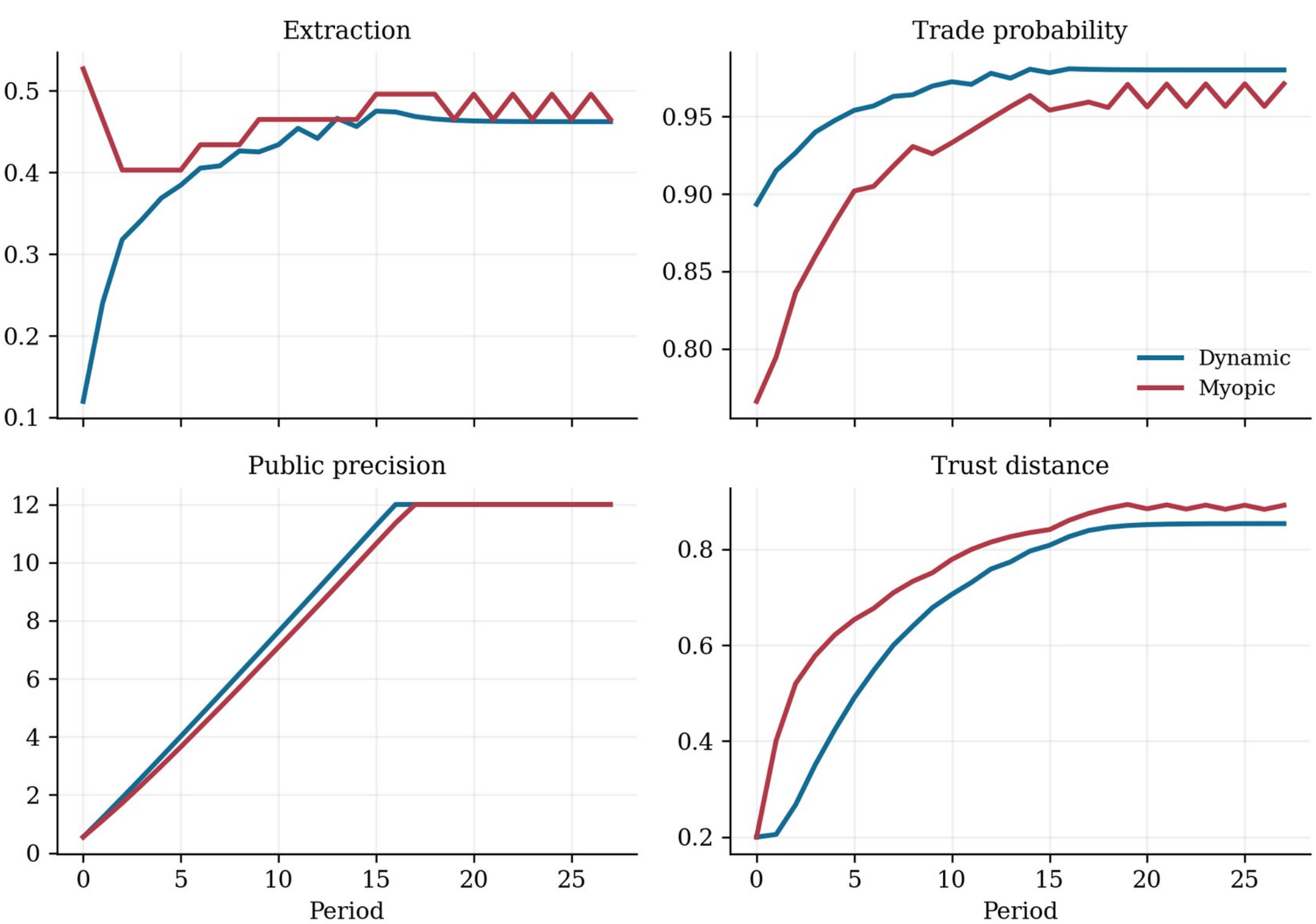

*Figure 3. Expected paths under dynamic and myopic policies.*

## 7. Selected Experience and Verifiable Updating

The symmetric benchmark deliberately removes selection on common quality: $m^S - m^B$ is independent of $\theta$. Real pricing and buying systems rarely have equal precision. When $\tau_S \neq \tau_B$, the more informative posterior loads more strongly on $\theta$, and the trading event contains information about common quality.

Write $w_j = \tau_j / (\tau + \tau_j)$ and $G = m^S - m^B$. Conditional on the public state, $(\theta, G)$ is jointly normal with

$$Cov(\theta, G) = \frac{w_S - w_B}{\tau} \tag{21}$$

and

$$\sigma_G^2 = \frac{\left(w_S - w_B\right)^2}{\tau} + \frac{w_S^2}{\tau_S} + \frac{w_B^2}{\tau_B}. \tag{22}$$

Trade occurs when $G \leq h$. Let $k = \left(h - E\left[G\right]\right) / \sigma_G$ and $\lambda\left(k\right) = \phi\left(k\right) / \Phi\left(k\right)$ be the inverse Mills ratio for left truncation.

**Proposition 4. Selection-aware and selection-naive learning**

With unequal private-signal precision,

$$E\left[z - \mu / \text{trade}\right] = -\frac{Cov\left(\theta, G\right)}{\sigma_G} \lambda\left(k\right). \tag{23}$$

The expression is zero in the symmetric benchmark. Otherwise, its absolute value rises when extraction lowers $h$, holding the signal structure fixed. A platform that applies the unselected Gaussian likelihood to traded outcomes has a nonzero expected update error. A platform that conditions on the known trading rule removes this truncation error under the maintained model.

The direction of bias depends on relative signal precision. If the pricing agent's posterior loads more heavily on common quality, the event that its estimate does not exceed the buying agent's estimate by too much selects lower common-quality realizations. Reversing relative precision reverses the sign. The general conclusion is about selection, not inevitable upward bias.

This distinction disciplines the "confidently wrong" claim. Correct Bayesian learning from selected observations is slower but not biased merely because the sample is selected. Bias arises when the platform treats transaction outcomes as though they were an unselected sample. Because the naive filter still adds precision after every observed outcome, reported uncertainty can contract while estimation error persists.

Verification can help at a precise boundary. A transaction record can commit to the signal schema, the offer and acceptance policy versions, the selection event, and the updating code. A proof, trusted execution report, or independent audit can establish that the declared selection-aware filter was applied to the committed inputs. It cannot prove that the Gaussian family is true, that unrecorded confounders are absent, or that $z$ equals latent utility. The distinction is between verified execution and validated knowledge.

This boundary produces four evidence levels. **Ordinary disclosure** states which update the platform claims to use but leaves execution unverifiable. **Audit** samples inputs and recomputes selected updates after the fact. **Trusted execution** can bind a running binary and protected inputs, subject to hardware and attestation assumptions. **Cryptographic proof** can establish execution of a specified relation without revealing all inputs, subject to the encoded circuit and setup assumptions. These technologies differ in cost, latency, privacy, and coverage, but they share the same semantic limitation: none proves a proposition that was not encoded or measured.

Selection-aware verification can nevertheless have economic value. A buyer agent may rationally impose an additional acceptance wedge when it cannot tell whether the public posterior was computed from selected data correctly. Removing that wedge expands $h$, increases current trade, and accelerates production of later outcomes. The benefit is largest when selection is strong, the relationship horizon is long, and the outcome signal is precise. But attesting a misspecified filter can make a wrong posterior more credible. Verification and external validation are complements precisely because the former increases reliance on the latter's maintained assumptions.

## 8. Welfare and Model-Risk Extensions

Two extensions delimit the core mechanism; full statements and proofs are in the online appendix. First, a completed experience can benefit later buyers through better matching and less bilateral disagreement. Let $J_S$ be the seller's continuation information value and $J_B$ the later buyers' net value after privacy loss, discriminatory extraction, and model error. In a strictly concave two-period benchmark, the private objective includes $J_S$, whereas welfare includes $J_S + J_B$. If $J_B > 0$, the seller prices the knowledge-producing transaction too high; an outcome-contingent transfer of $\beta J_B$ implements the constrained social price. If $J_B < 0$, the sign reverses. The result identifies the object that must be measured rather than presuming that more data are socially beneficial.

Second, agreement is not accuracy. If the common outcome stream is $z_t = \theta + c + v_t$ but both agents omit $c$, increasing public precision drives $E\left[\left(m^S - m^B\right)^2\right]$ to zero while each agent's mean-squared error converges to $c^2$. An execution certificate leaves that floor unchanged because it verifies the declared computation, not the maintained model. Independent holdouts, outcome audits, and model pluralism are therefore complements to execution evidence.

These extensions remain secondary to the paper's dynamic pricing results. They show why the transaction's information value may exceed the seller's private value and why a verified bilateral consensus can still be wrong. Figure 4 illustrates selection error and common-model risk; neither panel is used to prove the core restraint or extraction–learning results.

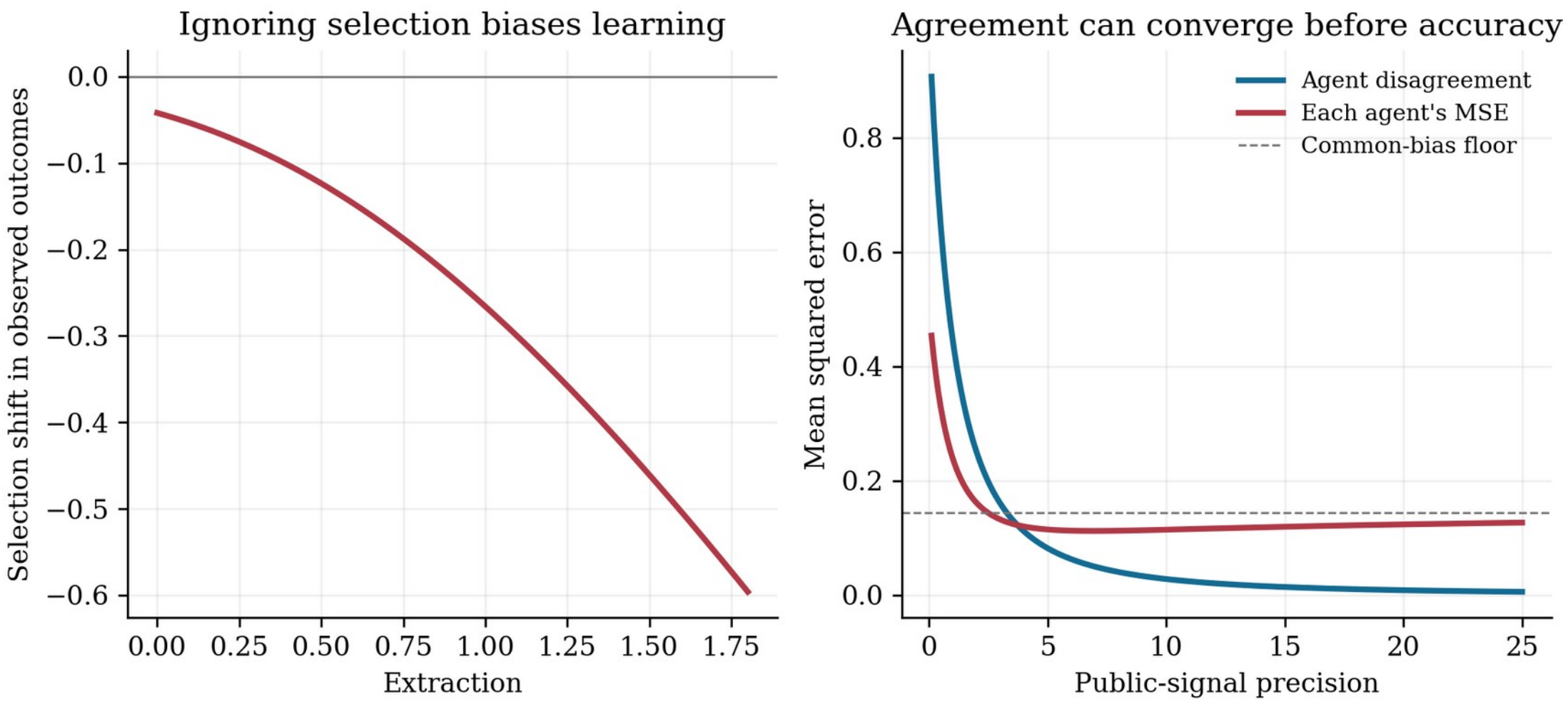

*Figure 4. Selection error and common-model risk.*

## 9. Numerical Policy Analysis

The numerical exercise solves the infinite-horizon Bellman equation rather than simulating an arbitrary heuristic. The baseline parameters are $\beta=0.92$, $\rho=0.74$, $\gamma=0.48$, $\chi=0.52$, $\ell-\underline{u}=1.20$, initial $M=1.10$, $\kappa=2$, and $\tau_z=0.75$. The grid contains 17 margin states on $[0.25, 1.95]$, 35 trust states on $[0, 2.4]$, 41 precision states on $[0.45, 12]$, and 51 actions on $[0, 1.55]$. Trilinear interpolation and five-node Gauss–Hermite quadrature evaluate continuation values. Value iteration converges in 248 iterations; an independent recomputation gives a sup-norm residual of $1.71\times10^{-9}$.

Three observations are robust to moderate changes in the calibration. First, the restrained region expands with trust distance. Second, dynamic extraction is below myopic extraction when knowledge is scarce, because the option value of producing the next outcome is largest there. Third, public precision does not have a globally monotone effect on extraction. It initially makes preserving transaction flow valuable; once the public state is precise, the marginal knowledge cost of a rejected trade falls. This nonmonotonicity is exactly why a model with only trust or only learning is incomplete.

The figures are transparent model computations, not empirical estimates. Every source array, parameter, interpolation rule, and convergence check is generated by the accompanying numerics.py. The online appendix reports robustness to the discount factor, signal precision, trust persistence, and outcome precision.

## 10. Empirical Implications

The model is empirically demanding because it distinguishes a trade decision from the information created by trade. The detailed record schema, likelihood, exclusion restrictions, structural nests, and controlled-market protocol are therefore placed in Online Appendix E. The main text retains the six predictions that discriminate the mechanism from generic demand learning.

The model yields six preregistrable predictions.

**P1: precision polarization.** Increasing public-signal precision raises trade when measured policy slack is positive and lowers trade when it is negative. The interaction, not the average precision effect, is the primary test of Proposition 1.

**P2: knowledge-sensitive restraint.** Holding current demand primitives fixed, a longer horizon and a more precise trade-contingent outcome reduce extraction most strongly when current public precision is low.

**P3: trust–knowledge complementarity.** Persistent identity magnifies the effect of outcome eligibility on restraint because the seller values the larger future transaction base created by both trust and learning.

**P4: selection gradient.** With unequal seller and buyer signal precision, the mean traded outcome changes systematically with extraction even conditional on the latent-state distribution. The sign reverses when the precision advantage reverses.

**P5: verifiable correction.** Attestation of a selection-aware update improves calibration and reduces a buyer suspicion wedge when the maintained model is correct. Attestation without external validation need not improve calibration under common misspecification.

**P6: agreement–accuracy divergence.** A shared biased public stream reduces seller–buyer disagreement faster than it reduces error against an external ground-truth measure.

These predictions distinguish the paper from a generic claim that more data improve pricing. P1 can have either sign; P4 can reverse with relative precision; and P6 predicts that an apparently desirable operational metric can improve while accuracy does not.

## 11. Managerial and Institutional Implications

### 11.1 Optimize the knowledge pipeline, not the transaction count

A firm that evaluates pricing only by current conversion and margin omits two assets: relationship capital and the future value of outcomes. Conversely, counting every completed transaction as an equally valuable data point is also wrong. Outcome quality, selection, consent, and relevance to future decisions determine whether a record improves knowledge.

The operational dashboard should therefore separate current revenue, trade probability, eligible outcome production, posterior calibration, and future trust. A policy can raise revenue while degrading calibration, or raise the data count while increasing selection bias. The model supplies a reason to measure these outcomes jointly.

### 11.2 Treat learning permissions as part of the contract

Authorization to purchase is not automatically authorization to pool consumption outcomes across principals, retain them indefinitely, retrain a pricing model, or redeploy the learned representation in future offers. A machine-readable mandate should distinguish collection, retention, aggregation, updating, and reuse. This is particularly important when one buyer's experience changes the prices or recommendations faced by others.

### 11.3 Verify the inferential path, not just the final score

A score alone does not reveal whether the platform corrected for transaction selection or used an eligible outcome. Useful evidence commits to input provenance, selection events, model version, update rule, and output. Yet even a perfect trace proves only that a specified computation occurred. External validity and common-bias testing remain separate governance functions.

### 11.4 Preserve independent information on both sides

Buying agents can discipline pricing agents only when their information or objectives are meaningfully independent. If both call the same foundation model, valuation API, or platform posterior, apparent bilateral confirmation can mask common-mode failure. Architecture choices should preserve buyer-private measurements or independent outcome audits when the cost of common error is material.

### 11.5 Where the mechanism is economically material

The model is most relevant when four conditions hold together. First, match value is uncertain before use. Second, consumption generates a measurable outcome that is more informative than the purchase decision alone. Third, the outcome can improve decisions beyond the current transaction. Fourth, offers remain attributable long enough for trust to affect later exchange. These conditions are not limited to consumer retail.

In business software, a buyer agent may evaluate a vendor's predicted productivity gain while the pricing agent estimates willingness to pay from firm characteristics. Deployment produces a realized workflow outcome that can update the vendor's cross-customer model. A high initial price can prevent the pilot needed to learn fit, and a failed or exploitative quote can influence later procurement. In cloud services, the corresponding outcome can be realized workload performance or cost. In professional marketplaces, it can be task quality. In insurance or credit, outcomes are highly informative but data rights, delayed realization, and distributive harm make $J_B$ especially ambiguous.

The mechanism is weaker for standardized search goods whose quality is already public, for one-shot anonymous transactions with no portable outcomes, and when consumption produces no reliable feedback. It is also weaker when nonpurchase produces an equally informative experiment–for example, when a platform can run a costless simulation of performance without deployment. In that case $\Delta_V$ in equation (19) approaches zero, although the relationship channel can remain.

The distinction helps managers decide whether to build a dynamic learning model at all. If the post-trade signal is weak, delayed beyond the relevant horizon, or unavailable for lawful aggregation, adding a nominal “data value” term to the pricing reward creates false precision. The empirical task is to estimate the incremental Blackwell value of the outcome conditional on information already available at the offer stage.

### 11.6 Limits of the analysis

The core model uses one seller and an exogenously chosen pair of delegated policy parameters. Competition can make outcome generation either a public-good problem or a proprietary-data investment, and endogenous principal choice can change the compatibility region. The online appendix outlines these extensions but does not solve their joint equilibrium.

Trust is summarized by one scalar state. Real counterparties can distinguish fairness, competence, privacy, and reliability, and a verifiable update may improve one dimension while worsening another. The scalar formulation is justified only when those responses can be projected onto an acceptance wedge. Empirical work should test that restriction rather than relabel every form of state dependence as trust.

Normal signals deliver transparent posterior disagreement and selection corrections. The main comparative statics extend to other distributions when public learning contracts disagreement and trade produces a more informative continuation experiment, but multimodal values or strategic reports can create additional equilibria. The paper also treats the outcome schema as fixed in the core. When firms choose what to measure, they may select metrics that improve extraction rather than match quality.

Finally, the theory does not make the planned software-agent benchmark an empirical estimate of consumer behavior. A language model that selects an action under explicit payoffs is evidence about policy execution, not about human fairness, market demand, or commercial trust. Those claims require different data. Maintaining this boundary is important because agent artifacts can fail arithmetic or instruction following in ways that are neither predicted consumer behavior nor an economic equilibrium response.

## 12. Discussion and Conclusion

The paper starts from a simple observation: for an experience good, a transaction can create an observation that no party possessed before consumption. Once price determines whether that observation exists, personalized pricing becomes an information-production problem. Once the offer affects a persistent trust state, it becomes a relationship-capital problem as well. Once seller and buyer agents possess different posteriors, common learning changes not only accuracy but the compatibility of delegated policies.

The model's strongest implication is conditional. More public knowledge makes compatible policies transact more reliably and incompatible policies fail more reliably. This knowledge-polarization result prevents a common mistake in discussions of AI markets: better prediction is not a substitute for institutional alignment. It makes the underlying mandates more binding.

Dynamic pricing then has a knowledge-flow term. Extraction can destroy the current experiment, raise future trust distance, and slow the resolution of disagreement. Correct selection-aware learning prevents statistical bias but not the data drought. Verifiable execution prevents neither misspecification nor a shared measurement bias. Finally, because later buyers benefit from some transaction-generated information, the private price of experimentation can be too high.

Simon framed management as intelligence, design, and choice under bounded rationality. AI agents lower the cost of executing and revising policies, but they do not eliminate the need to distinguish those stages. In this model, intelligence is endogenous to choice: the market learns only after the chosen policies allow trade. Review is therefore not an after-the-fact reporting function. It is the state transition that determines what the next decision can know.

The broader lesson is concise. Transactions produce data; institutions determine whether the data are produced; models determine what the data become; and verification determines only which parts of that path can be proven.

# Online Appendix to "When Trade Produces Knowledge: Dynamic Pricing, Bilateral Learning, and Trust"

**Chupeng Xie**

This appendix supplies posterior derivations, proofs, numerical details, robustness checks, additional identification arguments, and a preregistration-ready controlled-market design. Numbering prefixed by A refers to this appendix; unprefixed equations and propositions refer to the main text.

## A. Posterior Algebra and the Data-Generating Process

### A.1 Bilateral posterior means

Let the public belief be $\theta \sim N\left(\mu, \tau^{-1}\right)$, and let agent $j$ observe $q^j = \theta + \varepsilon_j$, where $\varepsilon_j \sim N\left(0, \tau_j^{-1}\right)$.

Normal conjugacy gives posterior precision $\tau + \tau_j$ and mean

$$m^j = \frac{\tau \mu + \tau_j q^j}{\tau + \tau_j} = \mu + w_j\left(\theta - \mu\right) + w_j \varepsilon_j, w_j = \frac{\tau_j}{\tau + \tau_j}. \tag{A1}$$

Hence the disagreement statistic $G = m^S - m^B$ is

$$G = \left(w_S - w_B\right)\left(\theta - \mu\right) + w_S \varepsilon_S - w_B \varepsilon_B. \tag{A2}$$

Its conditional mean is zero and its variance is

$$\sigma_G^2 = \frac{\left(w_S - w_B\right)^2}{\tau} + \frac{w_S^2}{\tau_S} + \frac{w_B^2}{\tau_B}. \tag{A3}$$

When $\tau_S = \tau_B = \kappa$, the first term vanishes and

$$\sigma_G^2 = \frac{2\kappa}{\left(\tau + \kappa\right)^2}. \tag{A4}$$

This is equation (9) in the main text.

### A.2 Why the symmetric benchmark is selection-free

In the symmetric benchmark,

$$G = \frac{\kappa}{\tau + \kappa}\left(\varepsilon_S - \varepsilon_B\right). \tag{A5}$$

Thus $G$ is independent of $\theta$ and of the outcome innovation $\nu$. The event $G \leq h$ is independent of $z = \theta + \nu$. Conditional on trade, the likelihood of $z$ is therefore the same Gaussian likelihood as its unconditional distribution. The update in main-text equation (12) is exact rather than an approximation.

This independence is a modeling benchmark, not an empirical presumption. Section 7 and Proposition 4 deliberately relax it. The separation is useful because it distinguishes two effects that are sometimes conflated:

1. **Data drought:** fewer trades produce fewer signals even under a correctly specified filter.
2. **Selection bias:** when trade depends on the latent state, an update that ignores the trading event is misspecified.

The first is present in the core dynamic model. The second requires asymmetric information or another source of correlation between acceptance and the latent state.

### A.3 Match-specific value

Suppose the current buyer's value is $v_i = \theta + \xi_i$, where $\xi_i \sim N(0, \tau_\xi^{-1})$. If both signals load on $v_i$, the same algebra applies with the prior variance of $v_i$ in the current transaction. An outcome may be decomposed as

$$z_i = \theta + \xi_i + \nu_i. \tag{A6}$$

If the platform observes many buyers and knows the variance components, it can update the common $\theta$ while treating $\xi_i + \nu_i$ as measurement noise. If the buyer also returns, its own history updates a relationship-specific posterior for $\xi_i$. The public state then becomes $(\mu_t, \tau_t)$ and the relationship state includes an individual posterior. The main text suppresses the latter to keep the trust and common-learning channels identifiable.

### A.4 Nonpurchase as information

The baseline assumes that the platform does not observe private signals, so rejection does not reveal $\theta$ under symmetric precision. If the platform observes the seller quote and knows the seller pricing rule, the quote may reveal $m^S$. If it also observes a rejection under a known buyer mandate, the event places an inequality restriction on $m^B$. A fully Bayesian platform should update from these events even without $z$.

Allowing such updates weakens but does not remove the trade-contingent mechanism. Consumption provides an additional experiment about realized match quality that an offer and rejection do not contain. Formally, let $E^0$ denote the experiment generated by offer and acceptance events, and let $E^1 = (E^0, z)$ denote the trade experiment. If $z$ is informative conditional on $E^0$, then $E^1$ Blackwell-dominates $E^0$. The knowledge-flow term in equation (18) becomes the continuation-value difference between these two experiments rather than between an update and no update.

## B. Proofs

### B.1 Proof of Proposition 1

Under symmetric signal precision,

$$D(a, r, \tau) = \Phi\left(\frac{h(a, r)}{\omega(\tau)}\right), \omega(\tau) = \frac{\sqrt{2\kappa}}{\tau + \kappa}. \quad \text{(A7)}$$

Since $\omega'(\tau) = -\sqrt{2\kappa} / (\tau + \kappa)^2 < 0$, differentiation yields

$$\frac{\partial D}{\partial \tau} = \phi\left(\frac{h}{\omega}\right)\left(-\frac{h\omega'}{\omega^2}\right). \quad \text{(A8)}$$

The normal density is positive and $-\omega' > 0$, so the sign equals the sign of $h$. As $\tau \to \infty$, $\omega(\tau) \to 0$. Therefore $h/\omega$ tends to positive infinity, zero, or negative infinity according as $h$ is positive, zero, or negative. Applying the corresponding limits of $\Phi$ proves the result. QED.

### B.2 Exact current profit and Bellman operator

In the symmetric benchmark, let $G = m^S - m^B$, $k = h/\omega$, and $M = \mu - \ell - c$. The pair $(m^S, G)$ is jointly normal and

$$Cov\left(m^S, G\right) = \frac{\kappa}{\left(\tau+\kappa\right)^2} = \frac{\omega^2}{2}. \quad \text{(A9)}$$

For a jointly normal pair, $E\left[m^S - \mu / G\right) = 1/2\, G$. The first lower partial moment of a centered normal variable is $E\left[G\, 1\{G \le h\}\right) = -\omega\,\phi(k)$. Therefore

$$E\left[\left(m^S - \ell + a - c\right) 1\{G \le h\}\right) = \left(M + a\right)\Phi(k) - \frac{\omega}{2}\phi(k). \quad \text{(A10)}$$

This proves main-text equation (15) and shows exactly why multiplying the expected margin by the acceptance probability is invalid.

Let $X = M \times R \times T$, where each factor is compact, let $A = [0, \acute{a}]$, and let $C(X)$ be the bounded continuous functions with the sup norm. If trade occurs, $M^+ = \Pi_M\left(M + \eta\right)$, where $\eta \sim N\left(0, \tau_z / [\tau(\tau + \tau_z)]\right)$. For $W \in C(X)$, first define the complete continuation operator

$$C_W(M, a, r, \tau) = D(a, r, \tau) E_\eta W\left(M^+, r', \tau^+\right) + \left[1 - D(a, r, \tau)\right) W(M, r', \tau).$$

The Bellman operator is

$$\left(T W\right)(M, r, \tau) = \max_{a \in A}\{\pi(M, a, r, \tau) + \beta C_W(M, a, r, \tau)\}. \quad \text{(A11)}$$

The transition is Feller: integration of a bounded continuous function against the continuous Gaussian kernel, followed by continuous projection, is continuous in the state and action. For any $W_1, W_2$, the maximum inequality and continuation weights imply

$$\| T W_1 - T W_2 \|_\infty \le \beta \| W_1 - W_2 \|_\infty. \quad \text{(A12)}$$

Thus $T$ is a contraction. The maximum theorem gives a nonempty, compact-valued, upper-hemicontinuous optimal-policy correspondence. Notice what this proof does not assert: neither higher precision nor a traded signal has globally positive private value. That sign depends on mandate compatibility, selected margins, and the continuation problem.

### B.3 Proof of Proposition 2

Existence, uniqueness, and existence of an optimal stationary policy follow from equation (A12), the contraction mapping theorem, and the maximum theorem.

Away from the caps and at differentiability points, write $D = D(a, r, \tau)$, $r' = \rho r + \gamma a$, $\acute{V}^{+} = E_{\eta} V(M + \eta, r', \tau^{+})$, and $V^{0} = V(M, r', \tau)$. The Bellman objective is

$$Q(a) = (M + a) D - \frac{\omega}{2} \phi(k) + \beta \left[ D \acute{V}^{+} + (1 - D) V^{0} \right]. \tag{A13}$$

Differentiating gives

$$Q_{a} = D + (M + a) D_{a} - \frac{k \phi(k)}{2} + \beta D_{a} \left( \acute{V}^{+} - V^{0} \right) + \beta \gamma \left[ D E_{\eta} V_{r}^{+} + (1 - D) V_{r}^{0} \right]. \tag{A14}$$

The third term is obtained from $\partial[-\omega \phi(k)/2]/\partial a = -k \phi(k)/2$. At any state where $\acute{V}^{+} - V^{0} \geq 0$ and the expected relationship derivatives are nonpositive, the last two terms are nonpositive because $D_{a} = -\phi(k)/\omega < 0$. The exact myopic derivative is the first three terms, not merely the usual margin–demand expression. Under strict concavity, adding a weakly negative marginal continuation term weakly lowers the unique maximizing action. At $a = 0$, concavity and $Q_{a}(0^{+}) \leq 0$ are sufficient for restraint. QED.

Strict concavity is sufficient rather than necessary. Gaussian acceptance need not make exact profit or the full state objective globally concave for every parameter vector. The numerical action grid therefore solves the global maximization and does not rely on the first-order condition.

### B.4 Proof of Proposition 3

Let the two systems begin from the same $(M_0, r_0, \tau_0)$. At period $t$, draw a common $U_t \sim Uniform[0,1]$. System $k \in \{A, R\}$ trades if $U_t \le D(a_t^k, r_t^k, \tau_t^k)$. When both trade, give them the same posterior-mean innovation. When only one trades, only that system updates $M$ and $\tau$. Assume inductively that $r_t^A \ge r_t^R$ and $\tau_t^A \le \tau_t^R$. The preserved policy ordering gives $a_t^A \ge a_t^R$. Because $D$ is decreasing in $a$ and $r$, and weakly increasing in $\tau$ on the maintained region,

$$D(a_t^A, r_t^A, \tau_t^A) \le D(a_t^R, r_t^R, \tau_t^R). \quad \text{(A15)}$$

The common-uniform construction implies that every trade on the aggressive path is also a trade on the restrained path. Hence cumulative transactions and precision increments are ordered. The trust transition is increasing in $r$ and $a$, so

$$r_{t+1}^A = min\{\acute{r}, \rho r_t^A + \gamma a_t^A\} \ge min\{\acute{r}, \rho r_t^R + \gamma a_t^R\} = r_{t+1}^R. \quad \text{(A16)}$$

This proves the induction for trust, precision, and cumulative trades. Posterior means need not be ordered when only the restrained system trades, which is why Proposition 3 imposes action ordering on realized coupled states and makes no claim about $M_t^A - M_t^R$. Conditional on a fixed current state and action, independent trials with probability $D$ have geometric waiting time $1/D$. The probability order gives the waiting-time conclusion. QED.

The policy-ordering condition rules out a pathological feedback in which the nominally aggressive policy becomes less extractive after reaching a worse state. A sufficient primitive implementation is a pair of fixed extraction rules or two monotone schedules separated by a nonnegative constant.

## B.5 Proof of Proposition 4

From equations (A1)–(A3), $(\theta, G)$ is jointly normal and

$$Cov(\theta, G) = Cov\left(\theta, (w_S - w_B)(\theta - \mu)\right) = \frac{w_S - w_B}{\tau}. \tag{A17}$$

For a jointly normal pair, the conditional mean is linear:

$$E[\theta - \mu / G] = \frac{Cov(\theta, G)}{\sigma_G^2}(G - E[G]). \tag{A18}$$

The mean of a normal variable left-truncated at $h$ is

$$E[G - E[G] / G \le h] = -\sigma_G \lambda(k), k = \frac{h - E[G]}{\sigma_G}. \tag{A19}$$

Iterated expectations and $E[v] = 0$ give

$$E[z - \mu / G \le h] = -\frac{Cov(\theta, G)}{\sigma_G}\lambda(k). \tag{A20}$$

When $w_S = w_B$, the covariance is zero. Otherwise the magnitude is proportional to $\lambda(k)$. The inverse Mills ratio for left truncation satisfies

$$\lambda'(k) = -\lambda(k)[k + \lambda(k)] < 0. \tag{A21}$$

Extraction reduces $h$ one for one and therefore reduces $k$, raising $\lambda(k)$ and the absolute selection shift. A likelihood that conditions on $G \le h$ divides the ordinary likelihood by the selection probability and restores the correct posterior under the maintained joint-normal model. QED.

## B.6 Welfare extension

The private objective has derivative

$$F_P(p) = D(p) + [p - c + \beta J_S] D'(p). \tag{A22}$$

The social objective has derivative

$$F_W(p) = D(p) + \left[p - c + \beta\left(J_S + J_B\right)\right] D'(p). \tag{A23}$$

At an interior private optimum $p^P$, $F_P\left(p^P\right) = 0$. Therefore

$$F_W\left(p^P\right) = \beta J_B D'\left(p^P\right). \tag{A24}$$

If $J_B > 0$ and $D' < 0$, this derivative is negative. Strict concavity of the social objective implies that its unique zero lies below $p^P$, so $p^W < p^P$. Adding a payment $s$ per completed eligible outcome changes the private continuation value to $J_S + s/\beta$. Setting $s = \beta J_B$ makes equations (A22) and (A23) identical. If $J_B < 0$, all inequalities reverse. QED.

### B.7 Common-model-risk extension

Let the public outcome average have precision $T$ and realization $\acute{z} = \theta + c + \acute{v}$, where $Var\left(\acute{v}\right) = T^{-1}$. Let agent $j$ combine it with any fixed finite-precision private source. Its posterior mean can be written

$$m_T^j = \alpha_{jT}\,\acute{z} + \left(1 - \alpha_{jT}\right)\widetilde{m}^j, \alpha_{jT} \to 1\left(j \in \{S, B\}\right). \tag{A25}$$

Because the public component is common,

$$m_T^S - m_T^B = \left(\alpha_{ST} - \alpha_{BT}\right)\acute{z} + \left(1 - \alpha_{ST}\right)\widetilde{m}^S - \left(1 - \alpha_{BT}\right)\widetilde{m}^B, \tag{A26}$$

whose mean squared value converges to zero when the private estimates have finite second moments. Meanwhile,

$$m_T^j - \theta = \alpha_{jT}\left(c + \acute{v}\right) + \left(1 - \alpha_{jT}\right)\left(\widetilde{m}^j - \theta\right). \tag{A27}$$

The noise variance tends to zero, the private term vanishes, and the expression converges in mean square to $c$. Thus mean squared error converges to $c^2$. An execution certificate changes none of these statistical limits because it does not change the maintained likelihood. QED.

## C. Numerical Implementation and Robustness

### C.1 Algorithm

The code creates a state-action tensor over public margin, trust distance, posterior precision, and extraction. Off-grid continuation values use trilinear interpolation. Five-node Gauss–Hermite quadrature integrates the posterior-mean innovation after trade. Value iteration begins at zero and stops at the declared sup-norm tolerance. After convergence, the program independently recomputes the exact Bellman right-hand side and records the fixed-point residual.

The baseline grid is:

| **Object** | **Range** | **Points** |
|---|---|---|
| Public margin $M$ | 0.25 to 1.95 | 17 |
| Trust distance $r$ | 0 to 2.40 | 35 |
| Public precision $\tau$ | 0.45 to 12.00 | 41 |
| Extraction $a$ | 0 to 1.55 | 51 |

The baseline converges in 248 iterations. Its independently recomputed sup-norm residual is $1.71 \times 10^{-9}$, and extraction is zero in 42.19 percent of grid states. These are diagnostics of the declared discretization, not estimates of a stationary distribution. numerical_verification.csv also records the exact-profit flag, posterior-mean grid size, and quadrature rule.

### C.2 One-dimensional parameter checks

The following table is regenerated on a coarser four-dimensional grid for computational transparency. Each row changes one primitive while holding the others at baseline. Exact entries are read from figures/policy_sensitivity.csv; they should not be copied from an earlier two-state operator.

| Parameter | Value | Restrained state share | Mean extraction |
|---|---|---|---|
| Discount factor $\beta$ | 0.80 | 0.413 | 0.238 |
| Discount factor $\beta$ | 0.97 | 0.470 | 0.211 |
| Trust persistence $\rho$ | 0.50 | 0.422 | 0.236 |
| Trust persistence $\rho$ | 0.90 | 0.493 | 0.198 |
| Trust response $\gamma$ | 0.25 | 0.417 | 0.239 |
| Trust response $\gamma$ | 0.75 | 0.480 | 0.205 |
| Outcome precision $\tau_z$ | 0.25 | 0.440 | 0.225 |
| Outcome precision $\tau_z$ | 1.50 | 0.457 | 0.218 |

Patience, trust persistence, antagonism, and outcome precision all expand restraint in this calibration. The outcome-precision effect is smaller because increasing $\tau_z$ makes each completed trade more valuable but also moves the system toward the precision cap more quickly. The paper does not claim global monotonicity outside the compatible region.

### C.3 Expected-path figure

Figure 3 is not a stochastic confidence band. It evaluates the dynamic and exact myopic policies from $(M_0, r_0, \tau_0) = (1.10, 0.20, 0.55)$ and uses

$$r_{t+1} = \rho r_t + \gamma a_t, \tau_{t+1} = min\{\acute{\tau}, \tau_t + \tau_z D(a_t, r_t, \tau_t)\} \tag{A28}$$

to plot the mean precision flow. Because the public posterior mean is a martingale under the maintained model, the plotted expected path holds $M$ at $M_0$. The Bellman solution does not: it integrates the traded posterior-mean innovation and discrete precision transition.

### C.4 Numerical integrity checks

The replication script performs or permits the following checks:

1. recomputation from a zero value function with no cached policy;
2. global maximization over the entire action grid rather than local first-order optimization;
3. post-convergence Bellman residual calculation;
4. export of every figure’s source values as CSV;
5. a separate coarser-grid sensitivity solve; and
6. deterministic plotting with no random simulation in reported figures.

## D. Extensions

### D.1 A buyer’s own learning

The core platform posterior is public. A repeat buyer can additionally maintain a private relationship posterior $(\mu^B_{it}, \tau^B_{it})$ from its own consumption. The buyer cap becomes

$$c^B_{it} = E[v_i / P_t, H^B_{it}] - \underline{u} - \chi r_{it}. \tag{A29}$$

The pricing agent may observe only an aggregate response rather than the private history. Bilateral disagreement then reflects public, seller-private, and buyer-private learning. The knowledge-polarization logic continues whenever the public update contracts the distribution of their difference, but the limiting disagreement need not vanish if buyer-private match information remains.

### D.2 Endogenous outcome quality

Consumption does not always generate a usable measurement. Let the seller choose measurement effort $e$ at cost $C(e)$, producing precision $\tau_z(e)$. Effort can include instrumenting the product, requesting structured feedback, or validating an operational result. The Bellman objective adds $-C(e)$ and replaces the fixed precision increment. The marginal condition is

$$C'(e)=\beta D(a,r,\tau)\frac{\partial E\left[V\left(M^{+}(e),r',\tau+\tau_z(e)\right)\right]}{\partial e}. \tag{A30}$$

Measurement effort and restrained pricing are complements when a higher trade probability raises the return to outcome quality. This extension also clarifies why a subsidy should condition on eligible information quality rather than on transaction count alone.

### D.3 Privacy and retention

Suppose a buyer bears privacy cost $k_d$ if its outcome enters the public update. Consent then requires compensation or an offsetting expected benefit. The net social information value becomes

$$J_B^{net}=J_B^{match}-k_d-J_B^{extraction}. \tag{A31}$$

The optimal instrument changes sign with this net value. A retention limit can be modeled as depreciation of public precision. If $\tau_{t+1}=\delta_\tau\tau_t+\tau_z y_t$, where $0<\delta_\tau<1$, the platform requires an ongoing flow of eligible outcomes to maintain knowledge. The extraction–learning feedback becomes stronger because a period without trade now loses precision in absolute terms.

### D.4 Competition

With competing sellers, one seller's low introductory price can generate knowledge that follows the buyer or becomes public. If the outcome is portable, rivals free ride on the experiment and private underproduction strengthens. If the outcome is proprietary, the first seller may subsidize trade to build a data advantage, as in data-enabled learning models. Trust can be seller-specific while public precision is shared, creating two separate dynamic externalities. A complete competitive equilibrium is beyond the paper's single-seller core and should not be inferred from the reduced-form welfare extension.

### D.5 Non-Gaussian learning

Normality supplies closed-form disagreement and selection expressions. The qualitative results require less. Knowledge polarization follows if public learning contracts the distribution of disagreement around zero in the sense that $G_\tau$ becomes a mean-preserving contraction. The extraction–learning mechanism requires only that higher extraction lower trade and that trade produce a Blackwell-more-informative continuation experiment. The shared-error result requires a common misspecified component whose weight dominates private information. Nonparametric or particle-filter implementations can preserve these features.

### D.6 Finite-horizon recursion and terminal behavior

Let $V_n(M, r, \tau)$ denote value with $n$ periods remaining and $V_0 = 0$. Writing $D = D(a, r, \tau)$, define the expected continuation value

$$C_{n-1}(M, a, r, \tau) = D\, E_\eta V_{n-1}(M^+, r', \tau^+) + (1 - D) V_{n-1}(M, r', \tau).$$

The finite-horizon recursion is therefore

$$V_n(M, r, \tau) = \max_a \{\pi(M, a, r, \tau) + \beta C_{n-1}(M, a, r, \tau)\}. \tag{A32}$$

In the final period, $V_1$ is the exact static objective and both dynamic terms disappear. For earlier periods, the derivative decomposition in equation (A14) holds with $V_{n-1}$. Whenever the continuation-value sign conditions in Proposition 2 hold, both dynamic terms are nonpositive; the paper does not impose that ordering globally.

This recursion yields a horizon effect that is directly testable in a controlled market. If the current state and primitives are held fixed, the extraction choice weakly approaches the static action as the remaining horizon contracts whenever the state objective has a unique maximum and the marginal continuation cost is ordered with horizon. The weak ordering can fail globally because public precision changes the value of future information, but the terminal convergence is exact. A benchmark that varies only a verbal instruction such as “think long term” does not implement this test; the policy must receive an economically different continuation payoff or remaining-horizon state.

**D.7 Endogenous principal mandates**

The core takes $\ell$ and $\underline{u}$ as institutionally chosen. To endogenize them, let the seller principal choose $\ell$ at cost $C_S(\ell)$, reflecting forgone current flexibility, and let the buyer principal choose $\underline{u}$ at protection cost $C_B(\underline{u})$, reflecting foregone matches. The induced slack is $h = \ell - \underline{u} - a - \chi r$.

If each principal ignores the other’s learning benefit, policy selection can be inefficient even when the period pricing agent solves its Bellman problem. The seller’s additional cushion increases trade and public information but transfers current surplus. The buyer’s stricter mandate protects current surplus but can suppress the experience needed to improve later matches. A platform that merely supplies a more accurate public posterior does not resolve this policy-selection externality; Proposition 1 makes the consequences of the selected slack more deterministic.

This extension connects the paper to bilateral delegation without duplicating a full principal-policy game. The main dynamic results condition on the selected policies. An empirical study can observe whether principals soften mandates when outcome eligibility, horizon, or verifiable updating raises the continuation value of trade.

## E. Empirical and Experimental Protocol

### E.1 Required records and identification

Estimation requires records that distinguish the trade decision from the information created by trade. A minimum record contains persistent but privacy-preserving seller, buyer, and relationship identifiers; pre-offer posterior summaries or committed score bins; pricing and acceptance policy versions; offer and response events; the post-consumption outcome, missingness reason, and timestamp; the public posterior before and after updating; later trust or switching proxies; and evidence coverage plus update-code version. Exact private estimates need not be revealed to the counterparty, but “no outcome because no trade occurred” cannot be pooled with ordinary missing measurement.

Randomized price or mandate slack identifies the causal effect of extraction on trade. Randomized attribution or identity persistence separates contemporaneous demand from the relationship transition. Randomized eligibility of an independently measured post-consumption signal identifies the value of public precision. Cross-randomizing seller- and buyer-signal precision identifies the selection term in equation (23), and randomizing verified execution while holding the signal fixed separates uncertainty about data use from signal content.

Let $y_t$ indicate trade and $o_t$ a valid observed outcome. A structural likelihood contribution is

$$L_t = D_t^{y_t}(1 - D_t)^{1 - y_t}\left[f(z_t / y_t = 1, I_t)\right]^{o_t} f(r_{t+1} / r_t, a_t), \tag{A33}$$

with a missingness model when $o_t < y_t$ and a truncation-corrected outcome density under unequal precision. The key exclusion restriction separates trust from information: an attributable price shock that contains no quality information shifts trust without changing the rational posterior, whereas an independently generated outcome shifts the posterior without changing the offer. Without such variation, declining demand can be fit by either lower expected value or higher trust distance.

### E.2 Confirmatory questions

A controlled study can preregister four primary questions:

1. Does increasing public precision raise trade under positive policy slack and lower it under negative slack?
2. Does a longer horizon or more persistent identity reduce seller extraction when outcomes are trade contingent?
3. Does unequal signal precision create the selection shift in equation (23), and does a selection-aware update remove it?
4. Does common public bias reduce agent disagreement faster than it reduces estimation error?

### E.3 Factorial design

The environment draws a latent common quality and independent seller and buyer signals. A full design crosses:

- policy slack: compatible, knife edge, incompatible;
- public precision: low, medium, high;
- relative signal precision: symmetric, seller-advantaged, buyer-advantaged;
- outcome availability: trade-contingent or externally observed;
- identity: resettable or persistent;
- horizon: one-shot or repeated;
- update evidence: opaque, disclosed, or execution-attested; and
- model specification: correct or common biased proxy.

Each software artifact receives a structured state and must return one permitted action. A pricing policy returns extraction or a price; a buying policy returns accept, reject, or an allowed review action. The economic environment, not the model, computes payoffs, trust transitions, outcome signals, and posterior truth. This prevents verbal explanations from being coded as economic behavior.

### E.4 Outcomes and exclusions

Primary outcomes are price/extraction, trade, eligible outcomes produced, posterior disagreement, posterior mean-squared error, and trust state. Secondary outcomes include invalid responses, constraint violations, redundant review, seller profit, buyer surplus, and total surplus net of measurement cost.

The preregistration should freeze model artifacts, prompts, task instances, random seeds, retry rules, structured-output schema, and analysis code before confirmatory collection. Technical failures should remain observable and be reported both as complete cases and under conservative bounds. Different sizes of one model family are scale checks, not independent families.

### E.5 Interpretation boundary

A local-model benchmark can establish that real software artifacts implement or fail to implement the declared policy mappings. It cannot estimate human trust, consumer fairness, field demand, or population treatment effects. Those claims require human subjects or commercial transactions. The theoretical propositions do not depend on LLM behavior; the benchmark is a construct-validation layer for the proposed agent implementation.

### E.6 Structural estimation details

Suppose the researcher observes public states, offers, mandate versions, trading outcomes, and post-trade measurements. If posterior summaries are observed, signal precision can be estimated from the mapping in equation (4). If only signals are experimentally assigned, the mapping is known by construction. If neither is available, seller and buyer information are generally not separately identified from offer and acceptance choices without additional exclusion restrictions.

Conditional on a candidate parameter vector

$$\Theta = \left(\kappa_S, \kappa_B, \tau_z, \ell, \underline{u}, \chi, \rho, \gamma, c, \beta\right), \tag{A34}$$

the researcher can solve the seller’s dynamic program and compute conditional choice probabilities after adding a small decision shock or measured implementation error. The buyer’s hard mandate gives the probability of acceptance from the disagreement distribution. The posterior transition supplies the outcome density, and equation (13) supplies the trust transition. Simulated maximum likelihood or Bayesian estimation can integrate over unobserved private signals.

Three problems require explicit treatment. First, policy endogeneity means that $a_t$ is correlated with $(M_t, r_t, \tau_t)$. The structural model uses this state dependence rather than treating price as exogenous; causal identification of demand still benefits from randomized price or an instrument that changes the seller's feasible action without changing match value. Second, attrition is state dependent. High trust distance can cause a principal to leave the platform, so a balanced surviving panel overrepresents successful relationships. Third, outcome validity can be endogenous even conditional on trade if satisfied buyers are more likely to report or if measurement fails selectively.

Model comparison should include at least four nests: static pricing; trust only; learning only; and trust plus learning. Out-of-sample predictions should be scored separately for trade, outcome calibration, and later relationship behavior. A model that improves current purchase prediction by using a latent trust state but fails outcome calibration has not established the paper's knowledge mechanism.

**E.7 Falsification and placebo tests**

The following findings would count against the proposed channels rather than being reinterpreted as support:

1. Public precision has the same effect on trade on both sides of measured policy slack, contradicting knowledge polarization.

2. Making an outcome available without changing any future decision has the same effect as making it decision-relevant, suggesting demand for the treatment label rather than information value.
3. Anonymous, nonattributable extraction changes later behavior as strongly as persistent attributed extraction, weakening the trust interpretation.
4. The selection-aware likelihood does not improve calibration under experimentally induced asymmetric precision and a correctly generated normal environment.
5. Agent disagreement remains a sufficient predictor of external error under a common biased signal treatment, contradicting the common-model-risk extension.
6. Dynamic pricing choices do not respond to horizon after the continuation payoff is made economically salient and the policy demonstrates payoff comprehension.

Placebo outcomes should include variables realized before the offer and outcome dimensions excluded from the public update. A claimed trust effect should not move pre-offer outcomes; a claimed knowledge effect should be strongest for dimensions measured by $z$, not for unrelated variables.

**E.8 Human-principal and field validation**

A human-principal experiment would ask participants to choose seller or buyer mandates while software agents execute the resulting policies. Participants would receive real monetary consequences over repeated rounds. This design tests delegation and willingness to preserve future learning, not whether humans can manually solve the Bellman equation. Treatments can vary whether principals see the knowledge-flow consequence of a rejected transaction and whether they can verify the platform update.

Field validation requires a setting with a meaningful post-use outcome and repeated or cross-user value. Candidate environments include software trials with verified usage outcomes, procurement pilots, service marketplaces, or platform-mediated tasks. The study must have a lawful basis for using outcome data and must not infer private valuations from sensitive data without appropriate governance. A credible design would randomize a price or fee component, outcome eligibility, and attribution while holding product access and measurement definitions stable.

Neither validation layer is required for the internal correctness of the theory. It changes the evidentiary claim. Software benchmarks speak to policy execution, human experiments to delegated preference and trust formation, and field studies to commercial magnitudes and external validity.

## F. Reproducibility Map

The replication directory contains:

- manuscript.md: main paper source;
- online-appendix.md: this appendix;
- numerics.py: Bellman solution, figures, CSV exports, and sensitivity checks;
- figures/fig1_knowledge_polarization.csv: Proposition 1 curves;
- figures/fig2_dynamic_policy.csv: complete policy surface;
- figures/fig3_policy_paths.csv: expected dynamic and myopic paths;
- figures/fig4_selection_and_common_error.csv: selection and common-error curves;
- figures/policy_sensitivity.csv: coarser-grid robustness results; and

- figures/numerical_verification.csv: convergence and residual checks; and
- figures/oracle_policy.npz: the full value and policy arrays used to construct the independent execution benchmark.

No empirical or human-subject data are used in this version. Numerical results are model implications. The experimental design in Section E is preregistration-ready but is not represented as collected evidence.